# An upper limit to the masses of stars

Donald F. Figer

*STScI, 3700 San Martin Drive, Baltimore, MD 21218*

**There is no accepted upper mass limit for stars. Such a basic quantity escapes both theory, because of incomplete understanding of star formation, and observation, because of incompleteness in surveying the Galaxy[1]. The Arches cluster[2,3,4,5,6,7] is ideal for such a test, being massive enough to expect stars at least as massive as 400 solar masses, and young enough for its most massive members to still be visible. I t is old enough to be free of its natal molecular cloud, and close enough, and at a well-established distance, for us to discern its individual stars[2]. Here I report an absence of stars with initial masses greater than 130 $M_\odot$ in the Arches cluster, where the typical mass function predicts 18. I conclude that this indicates a firm limit of 150 $M_\odot$ for stars as the probability that the observations are consistent with no limit is $10^{-8}$.**

Theory provides little guide in determining the most massive star that can form. Pulsational instabilities were once thought to destroy stars more massive than 95 $M_\odot$[8]; however, these pulsations may be damped[9]. Radiation pressure, and/or ionizing flux, inhibit accretion for stellar masses greater than 60 $M_\odot$[10], but direct collisions of protostellar clumps may overcome these effects[11]. While stellar evolution models have been computed for massive stars covering a large range in mass, up to 1000 $M_\odot$[12,13], no such stars have ever been observed. Indeed, some of the most massive candidates have proven to be systems of multiple stars[14].

Stars generally form with a frequency that decreases with increasing mass for masses greater than ~1 $M_\odot$, i.e. $d(\log N)/d(\log m) = G$, where $G$ is observed to be $-1.35$[15,16]. For stellar clusters young enough to not have lost members to supernovae, the distribution of stars is populated to the point where the mass function predicts one star, within the uncertainties of low number statistics. Therefore, stars with $M>150$ $M_\odot$ can only be observed in very massive clusters with total stellar mass $>10^4$ $M_\odot$. This requirement limits the potential sample of stellar clusters that can constrain the upper mass limit. Only a few clusters in the Galaxy satisfy this requirement, and all are located in the Galactic center.

To investigate the possibility that stars with $M>150$ $M_\odot$ exist, we obtained imaging data using the *Near-Infrared Camera and Multi-Object Spectrometer* instrument on the *Hubble Space Telescope* in a program to measure the mass functions of the most massive young clusters in the Galaxy, near the Galactic center[2]. Intervening dust prevents observations of these clusters at optical or ultraviolet wavelengths, so we obtained images in near-infrared wavelengths (see Supplementary Figure 1). We also imaged nearby control fields to estimate the number of field stars that contaminate our observations in such a densely populated region.

We extracted photometry for stellar sources in the images, and corrected for the absorbing effects of dust by comparing the observed colors to those expected for the appropriate spectral types; note that intrinsic colors of massive stars on the main sequence at infrared wavelengths differ by only a few percent. The dereddened fluxes were then converted into bolometric fluxes by accounting for the distance to the Galactic center. We then used the Geneva stellar evolution models to infer initial masses for each star[17] (see Figure 1). While these models have associated errors, note that the Arches stars are relatively unevolved; indeed, only the brightest dozen or so members show evidence of chemical enrichment by nucleosynthetic processes[18]. Some of the brightest stars in the cluster (three to ten, depending on cluster age within a range of 2 to 2.5 Myr and the coefficients in the extrapolation law) extend just above the 120 $M_\odot$ limit of the mass-flux relation; I estimate masses for them that do not exceed 130 $M_\odot$ through an extrapolation of this relation (see Supplmentary Figure 2).

The initial masses I estimate here agree with those inferred through wind/atmosphere modeling of high-resolution spectral observations to within a few percent[3]. Others have also applied the same technique to construct mass functions from infrared observations of massive young clusters, showing that these determinations are consistent with those estimated from optical observations[19]. In addition, several groups find good consistency in physical properties inferred from optical and infrared analyses for massive stars at all stages of evolution[20,21,22].

Figure 2 shows the resultant inital mass function of the Arches cluster, assuming an age of 2 Myr, for stars within a projected radius of 0.5 pc, and solar metallicity[2,18]. While the cluster is the densest in the Galaxy[3], the data do not suffer from incompleteness due to crowding or sensitivity for the four highest mass bins in the figure. The small amount of background contamination was removed by subtracting the number of stars observed in nearby fields; this resulted in subtracting a total of seven stars from the upper four populated mass bins. The frequency distribution generally decreases with increasing mass and is fit by two lines through the four most massive populated bins, which contain 39 stars. One line has a slope of $\Gamma=-0.9$, appropriate for the most recent determinations[2,23], and the other has a slope equal to the Salpeter value that is observed for most clusters. For both slopes, there appears to be a deficit of expected very massive stars with masses beyond ~130 $M_\odot$; variations in assumed age (±0.5 Myr), mass-loss rates and metallicity do not change the result. I estimate cumulative errors of ~10% and conclude conservatively that there is an upper mass cutoff of ~150 $M_\odot$ (see Supplementary Figure 3 for the effects of mass-loss on the most massive stars).

The observed deficit of stars in is significant. If there is no upper mass cutoff, then the odds of identifying no stars beyond the observed limit are $10^{-8}$ if 18 are expected, and $10^{-14}$ if 33 are expected, assuming Poisson statistics. In addition, the maximum predicted stellar mass is at least ~500-1100 $M_\odot$, values that are far beyond the masses inferred from the observations. I performed a Monte-Carlo simulation of model systems to predict the probability that a cluster with the mass of the Arches cluster could have no stars with initial masses greater than 130 $M_\odot$ as a function of cutoff mass (see Supplementary Figure 4). In this simulation, I added uncertainties due to differential extinction, photometric error, average cluster age, a spread of ages for individual stars, and error in estimating the average cluster age. Supplementary Figure 4 shows that the simulation predicts few systems with no stars having initial masses greater than 130 $M_\odot$ for cutoffs of 150 $M_\odot$ or greater.

Clearly, the cluster age is an important quantity for the analysis. If the cluster is too old, $t>3$ Myr, then its most massive members would no longer be visible, i.e. they would have progressed to supernovae, and the observations would then simply reveal an apparent cutoff due to the natural effects of stellar evolution. If the cluster is too young,

*t*~1 Myr, then the models would predict much higher initial masses for the brightest members; however, note that even younger ages would still require a firm upper mass cutoff, albeit at somewhat higher masses than predicted by the best estimated age. Analyses indicate that the cluster has an age of 2-2.5 Myr[2,3,18]. A younger age is inconsistent with the nitrogen-enriched atmospheres revealed in the spectra of the most massive stars in the cluster[18]. The fairly narrow age range is required by the observed heavy nitrogen enrichment in the brightest stars with relatively weak observed nitrogen content in the atmospheres of slightly lower mass stars[3,18]. An older age is inconsistent with the evolutionary status of the most massive stars in the cluster, i.e. they have not evolved to advanced stages of evolution, such as the carbon Wolf-Rayet phase[3,6]. In addition, the lack of any supernova remnants in the cluster argues for an age less than 3 Myr. Indeed, if massive stars filling the apparent deficit were formed and evolved to supernovae, one would expect that a supernova remnant would be formed at least every 50,000 years for the past 0.5 Myr, yet none are observed. In summary, stars with masses above ~150 $M_\odot$ should still be visible if they were formed, given our estimate of the age for the Arches cluster.

The observed upper mass limit is on the low side of the estimated masses of a few massive stars in the Galaxy, although it still falls within the error bars of these estimates. It is important to note the large errors in such estimates. For instance, many of these estimates rely on stellar wind/atmosphere models that do not model the effects of increased opacity produced by metals in stellar winds, i.e. line-blanketing. With more modern models, new mass estimates are less by up to a factor of two. In addition, mass estimates often suffer from uncertainties in distance, reddening, and photometry. The typical build-up of errors can easily result in an uncertainty of a factor of two in flux, and a similar factor in mass estimate. As an example, consider Pismis 24-1, which is estimated to have a mass of 210-290 $M_\odot$[24]. The build-up in errors for this star, from effects described above, produces at least a factor of two variation in flux estimates, and the original mass estimates were produced without the use of line blanketing. Once these combined effects are included, the true mass of this star may well be below 100 $M_\odot$. Note that uncertainty in distance is the next culprit in making accurate mass estimates once line-blanketing is included; however, the distance to the Galactic center is very well known to within 6%, and the Arches cluster is physically connected to phenomena known to be produced in the Galactic center[3].

If there are stellar systems more massive than the limit, then perhaps they are binaries, or products of mergers of lower mass stars. Indeed, the Pistol star, with an inferred initial mass of ~150-250 $M_\odot$[13], is surrounded by Wolf-Rayet and red supergiant stars that are older than the expected lifetime of such a star[25]. This apparent paradox may be reconciled if the star is actually multiple, or if it has recently experienced a rejuvenation through a merger with another star[26]. High spatial resolution imaging suggests that the Pistol star is not binary to within a limit of 110 AU[13], yet massive binaries can have components with orbits on yet smaller scales[14].

A cutoff of ~150 $M_\odot$ was found for R136 in the low metallicity environment of the nearby galaxy, the Large Magellanic Cloud (LMC)[27]. This result relies on an apparent deficit of 10 stars with masses beyond this limit, based on the assumption that R136 has a total stellar mass of $5(10^4)$ $M_\odot$; however, this high cluster mass includes stars that span a range of ages, including those that exceed the expected age when a massive star evolves to become a supernova. This has the effect of increasing the base of lower mass stars from which to extrapolate an expected number of higher mass stars, thus inflating an apparent deficit if those stars are not seen. Using a lower estimate of the cluster mass, $2(10^4)$ $M_\odot$[28], in stars sufficiently young for the present analysis, I estimate that the true deficit beyond 150 $M_\odot$ in R136 is roughly four stars, i.e. the result in the present work is more statistically significant by this measure. If the deficit of massive stars in R136 is real, then it represents another measurement of the upper mass cutoff.

Surprisingly, the cutoff may be similar in environments that span a factor of three in metallicity[18,29,30], although metal content is often cited as a proxy for the source of opacity that limits the infall of material and eventual build-up of massive stars. This result implies that the process that limits the mass of a star is independent of metallicity, at least in the range of metallicities primarily found within the Galaxy and the nearby LMC.

**Correspondence** and requests for materials should be addressed to D. Figer. (figer@stsci.edu).

**Acknowledgements**  I acknowledge useful discussions with Paco Najarro, Richard Larson, Nolan Walborn, Joachim Puls, Nino Panagia, Mark Morris, Carsten Weidner, Pavel Kroupa, Volker Bromm, R. Michael Rich, and Mario Livio.

**Competing interests statement**  The author declares that he has no competing financial interests.

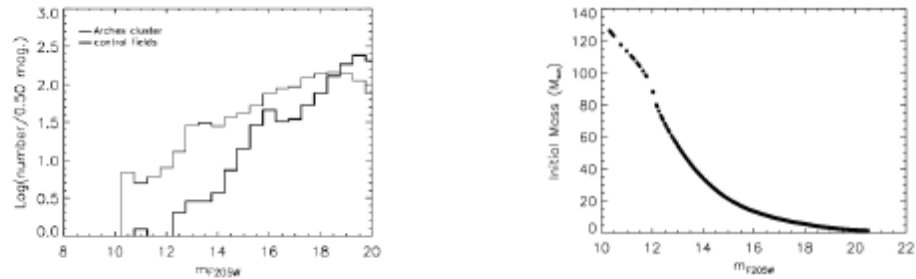

**Figure 1.** Observed frequency distribution and inferred masses of stars in the Arches cluster versus brightness. (left) Near-infrared ($\lambda_{center}$=2.05 μm) luminosity functions of the central parsec of the Arches cluster (thin) and nearby background fields (thick). There are generally fewer bright than faint stars in both fields; however for the vast majority of the brightness range, there are more stars in the Arches cluster than in the control fields. This allows an accurate subtraction of background stars in order to create a mass function for the cluster. The shapes of the distributions are consistent with a very young stellar cluster in the cluster field and an old population (>Gyr) in the control fields. (right) Inferred initial masses for Arches stars, based upon the Geneva models[17] for solar abundances and an age of 2 Myr. Each point represents one star in the cluster field. The three brightest stars have masses that slightly exceed 120 $M_\odot$, the upper limit of the models, and are assigned masses through a linear extrapolation of the mass-flux relation from points immediately below this value.

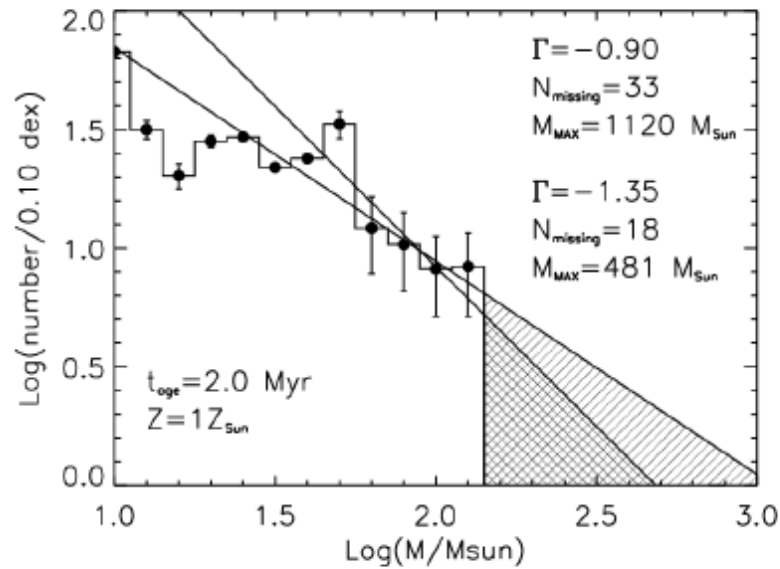

**Figure 2.** Frequency distribution versus mass for stars in the Arches cluster extracted from data in . The counts in each bin have been reduced by counts in nearby background fields. Error bars represent the Poisson errors based on the background subtracted counts. Two lines are drawn through the average counts in the four highest populated mass bins, with slopes inferred from the data $(d(\log N)/d(\log m)=\Gamma=-0.9)^2$ and that of Salpeter $(\Gamma=-1.35)^{15}$. For both lines, there is a clear deficit of stars with initial masses greater than ~150 M$_\odot$, as seen in the crosshatched regions. In addition, both slopes predict that at least one star in the cluster should have a mass far beyond that observed if there is no upper mass cutoff.